\documentstyle[aps,prb,multicol,psfig]{revtex}

\begin{document}
\title{Morphology of ledge patterns during step flow growth of
metal surfaces vicinal to fcc(001)}
\author{M. Rusanen$^{1,2}$, I. T. Koponen$^{1}$, T. Ala-Nissila$^{2}$,
C. Ghosh$^{3}$, and T. S. Rahman$^{3,2}$}
\address{$^{1}$Department of Physical Sciences, University of Helsinki, 
P.O. Box 64,\\
FIN--00014 University of Helsinki, Finland\\
$^{2}$Helsinki Institute of Physics and Laboratory of Physics,\\
Helsinki University of Technology,\\
P.O. Box 1100, FIN--02015 HUT, Espoo, Finland\\
$^{3}$Department of Physics, Kansas State University, Manhattan, KS 66506}
\maketitle

\begin{abstract}
The morphological development of step edge patterns in the presence of
meandering instability during step flow growth is studied by simulations and
numerical integration of a continuum model. It is demonstrated that the kink
Ehrlich-Schwoebel barrier responsible for the instability leads to an
invariant shape of the step profiles. The step morphologies 
change with increasing
coverage from a somewhat triangular shape to a more flat, 
invariant steady state form. The
average pattern shape extracted from the simulations is shown to be in good
agreement with that obtained from numerical integration of the
continuum theory.
\end{abstract}

\date{today}

\begin{multicols}{2}

Epitaxial growth on vicinal surfaces is known to give rise to interesting
growth instabilities under suitable conditions, {\it e.g.} to step bunching,
mound formation and meandering of the step edges \cite{Jeo99}. The
meandering instability emerges when the interlayer mass transport from the
upper side of the step is reduced due to the Ehrlich-Schwoebel barrier
\cite{Ehr66} enhancing growth of protrusions at the step edges. This is now
known as the Bales-Zangwill instability (BZI) \cite{Bal90} which tends to
destabilize the ledge morphology due to terrace diffusion and asymmetric
interlayer crossing. There is no diffusion along the ledges in BZI. However,
recently it was found that in the case of 
$1+1$ dimensional growth there is
an analogous phenomenon due to the {\it kink Ehrlich-Schwoebel barrier} for
going around a kink site at the step edge. The corresponding kink
Ehrlich-Schwoebel effect (KESE) leads to growth of unstable structures at
the step edges with a dynamically selected wavelength \cite{Pie99}. The
ledge instabilities were originally found and reported experimentally on the
Cu(1,1,17) vicinal surface \cite{Sch97} but attributed to the BZI scenario.
More recent STM experiments on the Cu(1,1,17) surface proposed that the
formation of the regular patterns is due to the KESE \cite{Mar99}. Since
then theoretical studies of the meandering instability have shown that the
KESE indeed supersedes the BZI in the formation of the periodic patterns
\cite{Pie99,Rus01,Kal01} and eventually leads to an in-phase motion of the
step edge structures \cite{Rus01,Kal01}.

Instability and wavelength selection of the step edge patterns due to
the KESE have been studied within the framework of a continuum step model
\cite{Pie99}, the solid-on-solid (SOS) lattice model \cite{Kal01,Mur99},
and semi-realistic Monte Carlo (MC) simulations \cite{Rus01}. 
In particular, in the recent MC work \cite{Rus01} it was shown
that on vicinal Cu(1,1,$m$) surfaces the
observed instability is due to the KESE and the competing BZI is of no
importance in the length and time scales considered. The 
role of dimer nucleation in determining the selected wavelength
was confirmed, in good agreement with the
theoretical scaling relation \cite{Pol97} and more recent SOS simulations
\cite{Kal01}. In the MC simulations, there was evidence of phase locking
of the ledge structures at the largest coverages studied, but this was not
quantitatively confirmed.

Regarding the step morphologies, a triangular shape has been predicted to
occur for a strong KESE and a rounded, more flat shape for a weak
KESE \cite{Pie99}. However, 
the MC simulations of Ref. \cite{Rus01} indicate that in the case
of a strong KESE there is in fact 
an interesting shape transition from narrow, somewhat triangular
shapes in the initial stage of growth to more rounded patterns in the large
coverage regime. Moreover, the MC simulations of Ref. \cite{Rus01}
and SOS model results of Ref. \cite{Kal01} are in disagreement with
asymptotic evolution of the step profiles as predicted
by the continuum theories \cite{Pie98,Gil00}.
In this work we study the ledge morphologies of growing steps on
the vicinal Cu(1,1,$m$) surfaces in detail. In particular, we
study the onset of the in-phase growth and the phase-locking 
of the step profiles in
the presence of a strong KESE. Our results show that the ledge morphologies
assume an invariant shape due to an interplay between various mass
transport currents and phase-locking of the steps. 
We show how the ledge profiles from the MC simulations can be reproduced
by explicitly including the relevant mass transport currents on the
surface. On continuum level this indicates a delicate balance
between the various currents that determines and
stabilizes the invariant ledge shapes.

The model system used here is as in Ref. \cite{Rus01},
based on MC simulations of a lattice gas model with energetics from
the effective medium theory (for more details, see Refs. \cite{Rus01,Hei99}). 
Our MC method is efficient enough to simulate growth of Cu
up to ten monolayers (ML) under realistic temperature and flux conditions.
The temperature range explored here was $T=240-310$ K and the 
flux $F=3\times 10^{-3}-1.0$ ML/s. Thus the
ratio between

\psfig{file=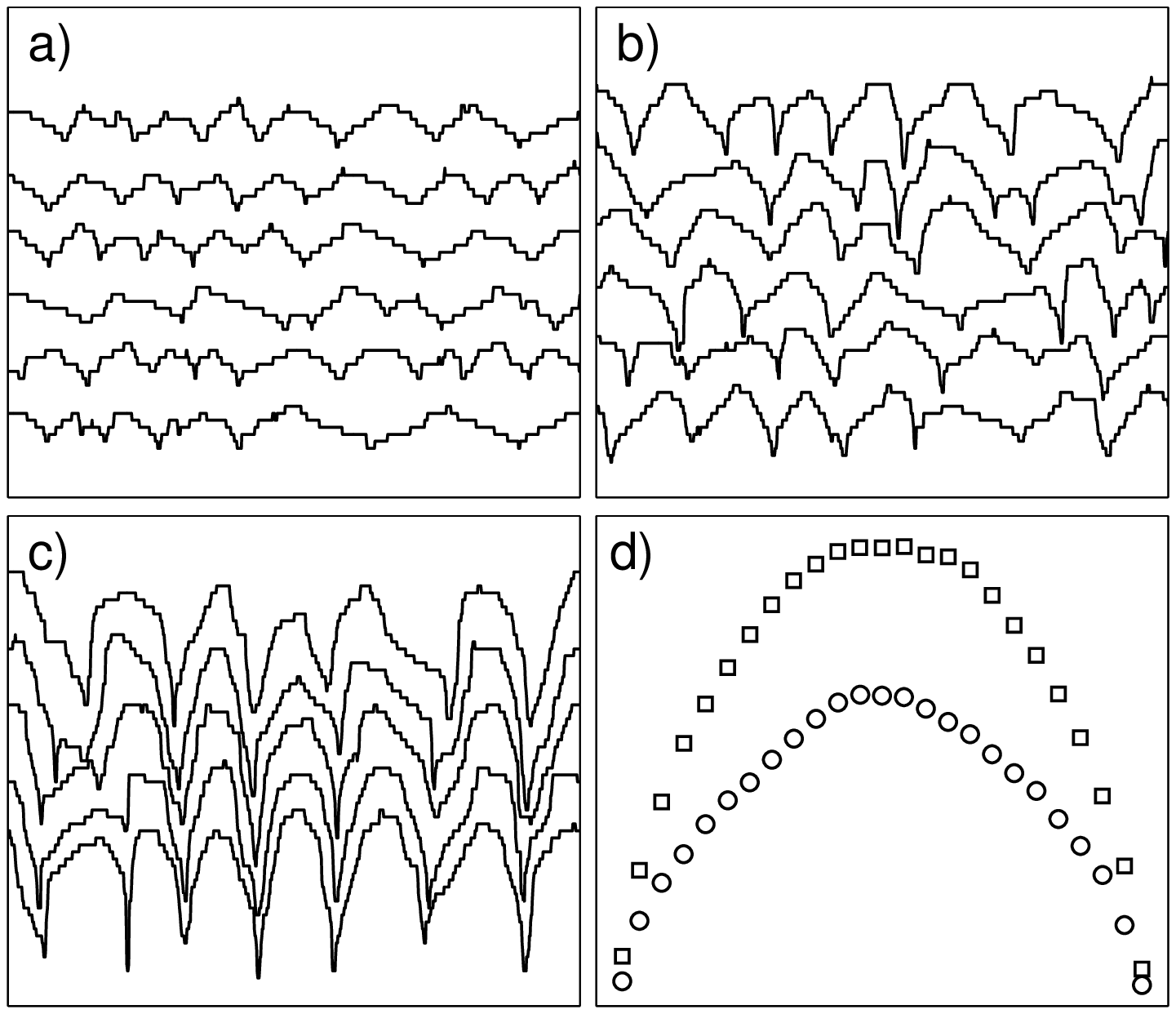,width=8.1cm}
\noindent {\bf Fig. 1:} Snapshots of typical ledge profiles with step
orientations in the close packed $\left[ 110\right]$ direction at $T=300$ K
with $F=8\times 10^{-2}$ ML/s for coverages $\theta =0.4, 2.0$, and
$10.0$, in figures (a)--(c) (lateral and vertical
scales are 1000 and 70 lattice spacings, respectively). In (d) the
shape transition is shown. The profiles have been obtained by averaging over
the meander periods at $\theta=0.4$ ML (circles) and at $\theta=2.0$ ML
(squares). The horizontal direction is scaled with the wavelength
$\lambda=120 a$ and in the vertical direction with the
roughness $w=1.4 a$ and $w=5.6 a$ for coverages $0.4$ ML and $2.0$ ML,
respectively.
\bigskip

\noindent the terrace diffusion and the flux
$D/F\approx 6\times 10^{5}-9\times 10^{7}$ in units of the lattice constant
$a=0.361$nm, corresponding to a typical molecular beam epitaxy
regime \cite{Kru97}. The energetics of the model also specifies the important
length scales controlling step flow growth. These are $\ell_{c}$, the
length scale for dimer nucleation at the step edge \cite{Pol97},
and the kink Schwoebel
length \cite{Pie99} $\ell_{s}=\exp [(E_{s}-E_{d})/k_{B}T]-1$ which is related
to the energy barriers $E_{s}=0.52$ eV and $E_{d}=0.26$ eV for jumps around a
kink site and along a straight edge, respectively. For the close packed
$\left[ 110\right] $ ledges, $\ell_{s}\simeq 10^{4}$ and $\ell_{c}\simeq
10^{2}$ around room temperature corresponding to
strong KESE \cite{Pie99,Kal01,Pie98}. 
In Ref. \cite{Rus01} it was shown that the wavelength of the step 
edge patterns is given by $\ell_{c}=(12D_{s}/FL)^{\alpha}$, $D_{s}$ being the
adatom diffusion constant along the straight edge, with a scaling
exponent $\alpha \approx 0.23$, and an effective barrier of
$E_{{\rm eff}}=75\pm 10$ meV. Both
are in good agreement with the exact values which give
$\alpha=1/4$ \cite{Pol97} and $E_{{\rm eff}}=E_{d}/4=65$ meV, respectively.
Our previous
study \cite{Rus01} was done on a Cu(1,1,17) surface but we have checked the
results also with smaller terrace widths.

Simulation results for the step edge profiles on Cu(1,1,17) are shown in
Figs. 1(a)--1(c) after deposition of $\theta =0.4,2.0$, and $10.0$ ML,
respectively, at $T=300$ K with $F=6\times 10^{-2}$ ML/s. In the beginning
of growth (Fig. 1(a)) the shape of the patterns is somewhat triangular as 
predicted for a relatively strong KESE \cite{Pie98}. The

\psfig{file=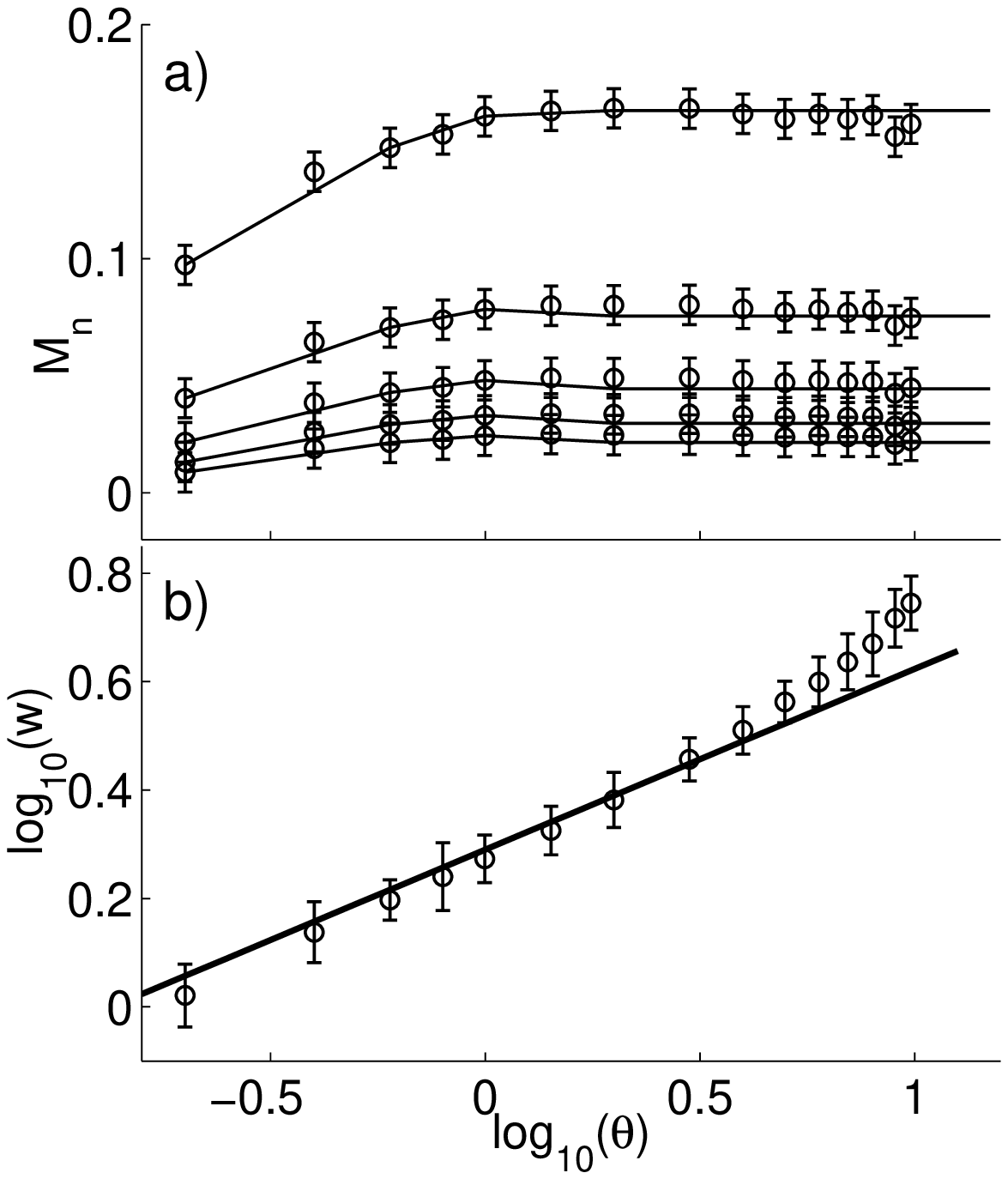,width=7.8cm}
\noindent {\bf Fig. 2:} (a) Lateral moments $M_{n}$ of
the step meander periods ($n=2,4,6,8,10$ from top to bottom) as a function
of coverage in a semi-logarithmic scale. The moments
approach constant values already before $2.0$ ML. (b) 
Width $w$ of the profiles as a function of the coverage. No
saturation is observed up to $10$ ML. The
slope of the solid line corresponds to $\beta = 1/3$.
\bigskip

\noindent  meandering structures
are not yet completely in the same phase indicating that the diffusion field
has not yet coupled the subsequent step edge trains, a typical feature for KESE
dominated meandering \cite{Kal01}. However, a selection of the relatively
well-defined wavelength for all ledges is apparent already at this stage of
growth \cite{Rus01}. At larger coverages the meandering of steps begins
gradually to phase-lock, seen in Fig. 1(b), and in-phase growth
and phase-locking seem complete at largest studied coverage 
of $10$ ML shown
in Fig 1(c). However, now the average shape of the patterns is clearly
different from that at low coverages. The shape of the average
patterns is more rounded, as predicted for a weak KESE. 
In Fig. 1(d) we show this change by comparing average ledge
profiles after $0.4$ and $2.0$ ML, respectively \cite{profiles}.

From Fig. 1 it is clear that there is no coarsening of the structures when
the coverage is large enough. The steady state pattern shape seems to be
governed by geometric constraints which is a sign of asymmetry of the
growth rates between bottom and top parts of the steps.  This asymmetry
is a general feature in many models of step growth with or without
coarsening \cite{Pol96,Hei98}. Moreover, a
quantitative inspection of the patterns at larger coverages suggest that the
profiles have an invariant shape. This can be seen by examining the $n$th
lateral moments $M_n(\theta)=\langle \zeta_i(x,\theta) x^n \rangle_{i,x}$ of
the meander periods $\zeta_i(x,\theta)$, shown in Fig. 2(a).
The scaled even moments approach

\psfig{file=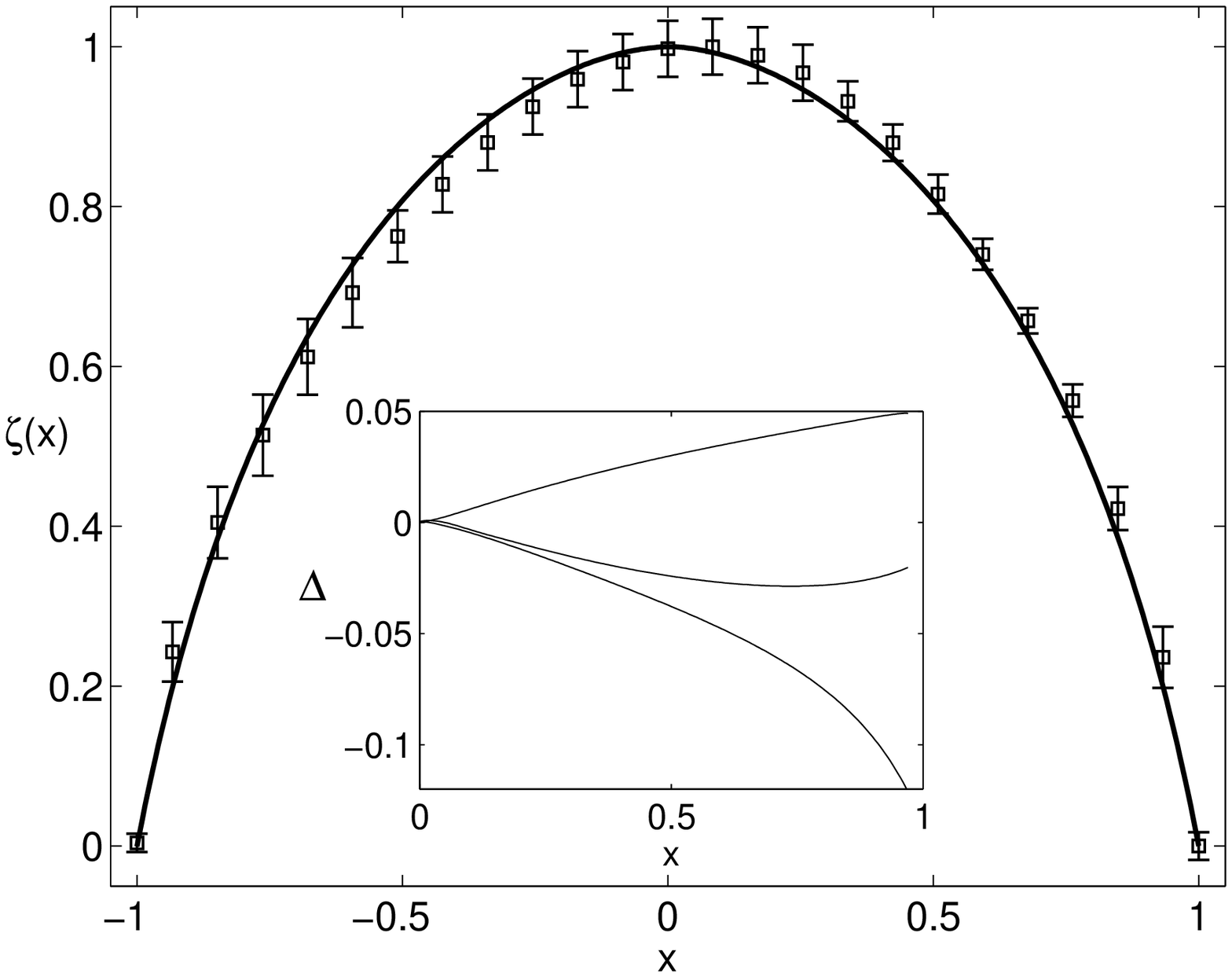,width=8.1cm}
\noindent {\bf Fig. 3:} The average shape of the step patterns at $T=300$ K
and $\theta=8.0$ ML with $F=6 \times 10^{-2}$ ML/s (squares).
Changing the
temperature or the flux as described in the text
does not have any effects within the error
bars. The solid line is the stationary profile obtained by integration of
Eq. (\ref{currents}). Good agreement between the average and the integrated
profiles is evident. The inset displays the relative differences of the
profiles, $\Delta=(\zeta_{all} - \zeta_{i})/\zeta_{all}$, where $\zeta_{all}$
is the profile with all currents included, and $i=k,e,SB$ denotes the solution
with only a small contribution for the KESE current from the Gibbs-Thomson
and symmetry breaking currents, respectively (from top to
bottom in the inset).
\bigskip

\noindent  their steady state
values already at $\theta \approx 2$ ML. The patterns are still
changing, however, which can be seen from the roughness of the step
$w(\theta)=\sqrt{\left\langle \zeta(x,\theta)^{2}\right\rangle_x}$, where
$\zeta(x,\theta)$ is the step profile. It does not show any sign of
saturation up to the largest coverage in the simulations. Instead the
roughness follows $w(\theta )\sim \theta^{\beta}$,
with $\beta \approx 0.3$ as shown in Fig. 2(b). It is interesting to note
that although the roughness does not saturate the shape
of the periodic structures attains an invariant form.

Our simulation results show that the profile shape is rather insensitive to
deposition and temperature conditions. This suggests that the invariant shape
is not dependent on the relative magnitudes of the various diffusion processes
but rather is a result of geometric constraints due to crowding and in-phase
evolution of the step edges. In order to justify this assumption we compare
the MC profiles with continuum profiles which are obtained as
stationary solutions to the dynamic equation
$\partial_{t}\zeta =-\partial_{x}J_{tot},$ where $J_{tot}$ is the total mass
current at the step edge.
The most important partial currents which we take into account in the total
current here, when expressed in terms of the variable
$m(x)=(\partial_{x}\zeta )/\sqrt{1+(\partial _{x}\zeta )^{2}}$ and
appropriately scaled, are the mass current due to the destabilizing
strong KESE \cite{Pie99}
\begin{equation}
\label{current:kese}
J_{k} = \frac{m(\sqrt{1-m^{2}}-\left| m\right| )\sqrt{1-m^{2}}}
	     {(\left|m\right| +L_{c}^{-1}\sqrt{1-m^{2}})^{2}},
\end{equation}
%

\psfig{file=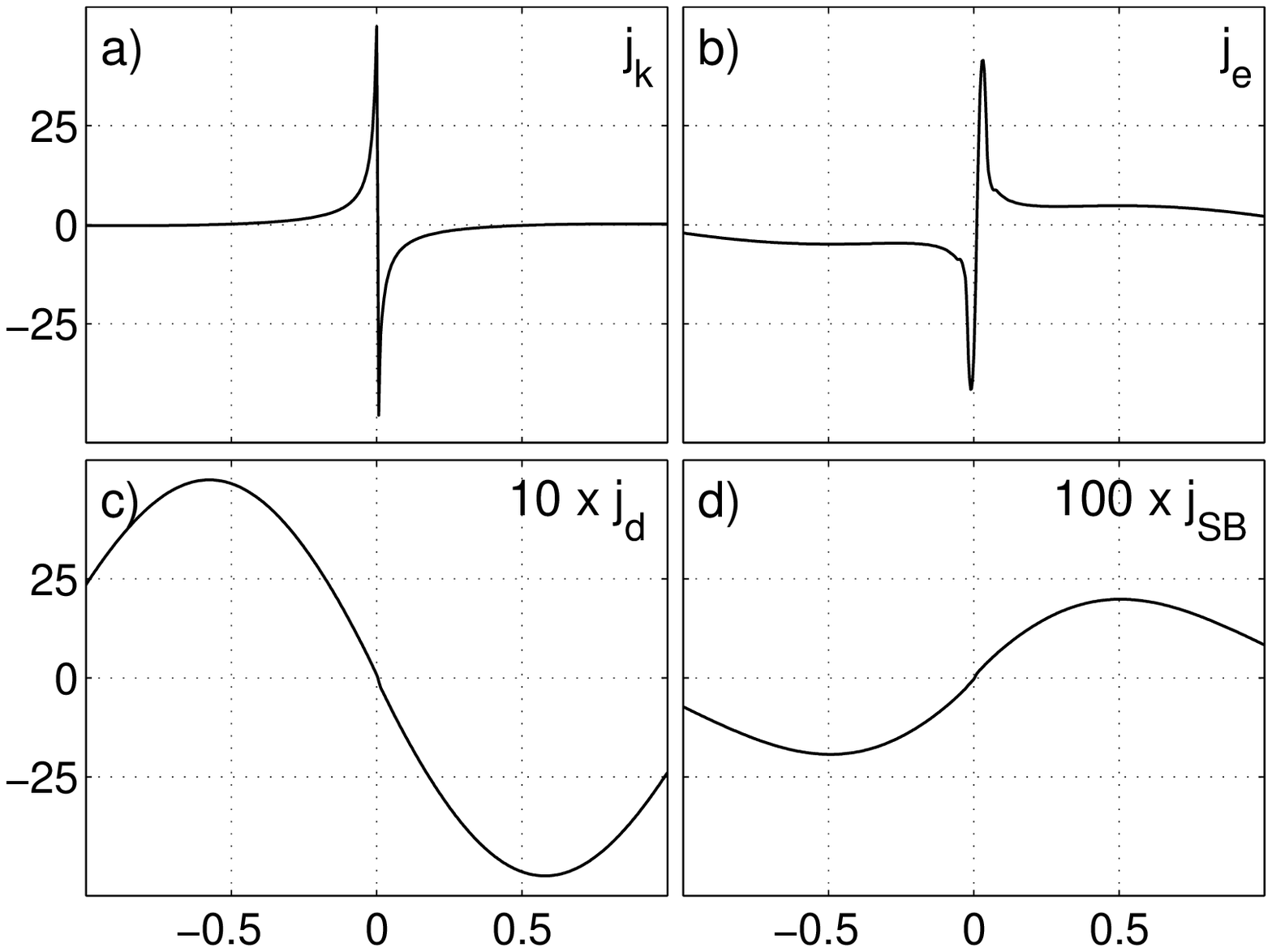,width=8.1cm}
\noindent {\bf Fig. 4:} The mass currents Eqs. (1)-(4) 
are shown using the integrated profile as an input. Note the difference
between the vertical scales for (c) and (d).
\bigskip

\noindent the stabilizing current due to the Gibbs-Thomson effect and edge
diffusion \cite{Pie98,Gil00,Kal00}
\begin{equation}
\label{current:gt}
J_{e} = \frac{2D_{S}\tilde{\Gamma}}{F}\left( \sqrt{1-m^{2}}
         + \frac{D_{L}}{D_{S}L}\right) (\partial_{xx}m)\sqrt{1-m^{2}},
\end{equation}
the current into the step edge from the deposition flux
\cite{Pie98,Gil00,Kal00}
\begin{equation}
\label{current:dep}
J_{d} = L m \sqrt{1-m^{2}},
\end{equation}
and the front-back symmetry breaking current \cite{Gil00,Pol96}
\begin{eqnarray}
\label{current:sb}
J_{SB} &=& -\frac{D_{S}\tilde{\Gamma}L}{F}(\partial_{x}m)(\partial_{xx}m)
   \sqrt{1-m^{2}} \nonumber \\
  & &+ \frac{L^{2}}{3}m(\partial_{x}m)(3-m^{2}).
\end{eqnarray}
In these expressions $L$ is the terrace width, $D_{S}$ is the macroscopic
diffusion constant on the terrace, $D_{L}$ is the macroscopic diffusion
constant along the step edge, and $\tilde{\Gamma}$ is the step stiffness (see
Refs. \cite{Gil00,Gie01} for the definitions and experimental values of the
parameters, respectively). All length scales are given in the units of the
lattice constant. By requiring the condition of stationarity
\begin{equation}
\label{currents}
J_{tot} \equiv J_{k}+J_{e}+J_{d}+J_{SB} = 0,
\end{equation}
we obtain a second order differential equation for $m(x)$.
The stationary profiles are obtained
by solving Eq. (\ref{currents}) numerically for given initial conditions
$m(\pm 1)=\pm m_{0}$.

The stationary solution is found using the value $m_{0}\approx 0.97$ as the
boundary condition in order to match the end points with the slopes of the
patterns obtained from the MC simulations \cite{m0}. The other parameter
values of the integration are based on known energetics of Cu, yielding
$\ell _{c}=700-1600$, $\ell _{s}=2\times 10^{4}-2\times 10^{5}$,
$D_{L}/(D_{S}L)=250-700$, and $D_{S}\tilde{\Gamma}/F=0.5-4000$ in the range
$T=240-300$K and $F=3\times 10^{-3}-10^{-1}$ ML/s. For the step
stiffness we used the expression $\tilde{\Gamma}=\exp {[E_{k}/k_{B}T]}/2$, where
$E_{k}=0.13$ eV is the kink energy \cite{Gie93}.
In all cases we set $L=10$ for the terrace
width. The resulting profiles are shown
in Fig. 3 with various values of the parameters. The shape is rather
independent of the details of the currents in agreement with simulations.
In Fig. 3 the average shapes obtained from the simulations are plotted with
a few different flux rates. In the inset we show how the resulting profile
deviates from the complete one when each of the mass currents is forced to be
small.

In Fig. 4 we show the mass currents using the integrated profile as an input.
It is now seen that for the invariant profile there is a delicate
compensation of the currents, the Gibbs-Thomson current compensated by the
sum of the KESE, the deposition, and the symmetry breaking currents. This
compensation happens for the specific shape of the profile, and cannot
take place {\it e.g.} in the case of a triangular shaped profile as obtained in the
initial stages of growth. In determination of the stationary profile shape
the front-back symmetry breaking and the geometric constraints contained
implicitly in the initial conditions are crucial.

In summary, the MC \cite{Rus01} and the SOS \cite{Kal01} simulations have
proven that the KESE is the dominant mechanism behind the meandering instability
and that it leads to the selection of the dominant wavelength determined by
dimer nucleation at step edges. In this work we have shown that the KESE
also induces an invariant shape of the step profiles during in-phase
growth. This occurs even though the overall roughness of the step
structures $w(\theta)$ shows no signs of saturation. The value of the 
corresponding scaling exponent
$\beta \approx 0.3$ is consistent with the case of an 
isolated step \cite{Sal93}. The
SOS model gives for the strong KESE an exponent $\beta \approx 0.57$
\cite{Pie99}, while for a collection of steps in the phase-locking regime
$\beta = 1/2$ \cite{Pie98,Gil00,Kal00}. This puzzling behavior of dynamical
scaling is apparently related to the strict in-phase growth and consequent
formation of the invariant shape of the profile. The fact that the shape
remains invariant although the roughness does not show any sign of saturation
indicates a subtle coupling of the step edge currents with the stationary
morphology. By numerically integrating the continuum equation we have 
shown how the interplay between various surface currents determines the
invariant step shapes.

Acknowledgements: We wish to thank J. Kallunki and Z. Chvoj for helpful
discussions and the Center of Scientific Computing, Ltd. 
for computing time. TSR and CG thank colleagues at Fyslab, HUT for their
wonderful hospitality during their stay in Helsinki. This work
has been supported by the Academy of Finland, in part through its
Center of Excellence program. We also acknowledge partial support from
the National Science Foundation, USA under Grant EEC-0085604 (including
International Supplement).

\end{multicols}

\begin{references}
\bibitem{profiles}
The profiles have been obtained by averaging over the
meander periods $\zeta_i(x,\theta)$ and scaling each
period as $x \rightarrow x/\lambda$,
$\zeta_i \rightarrow \zeta_i/w(\theta)$, where $x$ is the coordinate
between the end points of the meander period along the step direction,
$\lambda$ is the wavelength of the period, and $w(\theta)$ is the roughness.
\bibitem{method} To integrate Eq.~(\ref{currents}) we used the standard
Runge-Kutta method over one meander period and checked the accuracy of the
results by monitoring the local and global errors during the integration.
\bibitem{m0} The value of $m_0$ determines the behavior of the profile at
the cusp and is basically given by the simulation results. One could
also use an arbitrary choice $0 < m_0 < 1$ where $m_0 = 0$ implies
a flat shape and $m_0 = 1$ the diverging
profile at the cusp. However, this choice gives the same results
if the slopes with the MC profiles are matched.
\end{references}
\end{document}